\documentclass[mathleft]{an}
\usepackage{graphicx}
\usepackage{times}

\newcommand{\procspie}{Proc. SPIE}
\newcommand{\aap}{A\&A}

\begin{document}

\Pagespan{789}{}
\Yearpublication{2006}%
\Yearsubmission{2005}%
\Month{11}%
\Volume{999}%
\Issue{88}%

\title{X-ray Emission from GPS and CSS Sources}

\author{Aneta Siemiginowska\inst{1}\fnmsep\thanks{Corresponding author:
  \email{asiemiginowska@cfa.harvard.edu}\newline}
}
\titlerunning{X-ray Emission}
\authorrunning{Aneta Siemiginowska}
\institute{
Harvard-Smithsonian Center for Astrophysics, 60 Garden St., Cambridge, MA 02138, USA
}

\received{}
\accepted{}
\publonline{}


\abstract{Many X-ray observations of GigaHertz Peaked Spectrum and 
Compact Steep Spectrum sources have been made with Chandra X-ray
Observatory and XMM-Newton Observatory over the last few years. The
X-ray spectra contribute the important information to the total energy
distribution of the compact radio sources. In addition the spatial
resolution of Chandra allows for studies of the X-ray morphology of
these sources on arcsec scales and provide a direct view of their
environments.  This paper gives a review of the current status of the
X-ray observations and their contribution to our understanding of the
nature of these compact radio sources. It also describes primary
physical processes that lead to the observed X-ray emission and
summarize X-ray emission properties expected from interactions between
an expanding radio source and the intergalactic environment.
}
\maketitle

\section{Introduction}

A standard picture of a radio source shows a large scale emission in
forms of lobes and hot spots that are often connected to the radio
core by narrow jets. Such emission spans hundreds of kiloparsecs and
can be studied in details in nearby sources, for example in the
closest powerful FRII radio source Cygnus A (FR II, z =
0.056). However, the morphology of the X-ray emission is very
different.  The {\it Chandra} X-ray Observatory image of Cygnus A
(Wilson, Young \& Shopbell 2000, Smith et al. 2002) shows a more
circular diffuse X-ray emission surrounding the radio source with two
prominent hot spots and a filamentary structure.  A primary source of
this X-ray emission is a hot (a few keV) thermal cluster gas, while
the radio emission is mainly due to non-thermal particles accelerated
in the jet, lobes and hot spots.

The nearby (3.7~Mpc) radio source Centaurus A is much smaller than
Cygnus A. It has a double lobe radio structure only slightly larger
than its host galaxy.  The source morphology is very rich, showing an
absorbed core, dust lanes, jet, counter-jet, radio lobes and shocks in
multi-wavelength observations. X-ray emitting components can be
resolved in X-rays because of the source proximity. They provide
critical information about the physics and processes that are
happening in this source (e.g. Kraft et al. 2007).  GPS radio sources
are embedded within the host galaxy and their multi-wavelength
morphology might be similar to Cen A. However, if they are more
distant their X-ray components may not be resolved.

What type of processes can generate an X-ray emission?  Thermal
emission is associated with a hot plasma with temperatures of a few
keV and is observed in clusters of galaxies or halos of some active
galaxies. There have been some failed attempts to detect such emission
surrounding GPS sources in the past searches for a ``confinement
medium'' of the radio source (Antonelli \& Fiore 1997, O'Dea et al.
2000, Siemiginowska et al. 2003) in the past.  Non-thermal emission
associated with relativistic particles requires acceleration processes
and is observed in jets, or shock regions resulting from interactions
between the radio source and the ISM. Relativistic particles emit
radio synchrotron photons and radio observations can give information
about the distribution of particles. X-ray synchrotron emission requires
very high energy electrons ($\gamma \sim 10^7-10^8$) that have short
($\sim10$~years) synchrotron life times. Therefore any X-ray
synchrotron emission requires an on-going re-acceleration process.

Studies of interactions between the radio source and the interstellar
and intergalactic medium can provide information about the energy that
is deposited into the IGM by the expanding radio source. This is
important to our understanding of the feedback process. GPS/CSS
sources are observed at the early stage of their growth, where such
interactions are strong. In addition theoretical models predict that a
relatively strong X-ray emission should be associated with GPS
and CSS sources.

\section{X-ray Emission of Radio Sources: Models}

There are several theoretical predictions of the X-ray emission from
evolving radio sources.  Begelman \& Cioffi (1989) following Scheuer
(1974), draw a general picture of an \break evolving radio source
where the shocked IGM creates an overpressured cocoon surrounding a
double radio structure with lobes, jets and a bow shock bounding the
cocoon. The shock heated material emits X-rays. Reynolds, Heinz \&
\break Begelman (2001) present numerical simulations of a radio source
expansion within a uniform medium with three \break phases of the
evolution: a supersonic cocoon, subsonic sideways expansion with a
weak shock and supersonic jet, and a final sonic boom
phase. Each phase is characterized by a different morphology of the
X-ray emitting gas that is being heated by the expanding
source. Sutherland \& Bicknell (2007a, 2007b) shows a similar view of
the radio source evolution within a clumpy medium (see also Bicknell
\& Sutherland 2006). Also in this simulations the X-ray emission is
the result of interactions and the X-ray luminosity depends on the
density and clumpiness of the medium. They discuss the time dependent
(0.1-10~keV) X-ray emissivity with the maximum of the X-ray luminosity
reached within the first few $\sim 10^4$~years.

Heinz, Reynolds \& Begelman (1998) consider an evolution of GPS
sources within the uniform environment of the host galaxy. The
simulated X-ray surface brightness shows the shock discontinuity
moving outwards with time. X-ray detections of such discontinuity
require high spatial resolution and dynamic range observations in
nearby sources. The shock of Cen A is seen as a factor of $\sim10$
jump at $\sim7$~kpc away from the nucleus and it is $\sim600$~pc wide
(Kraft et al. 2007). It has not been detected in any observations of
GPS sources so far, possibly due to the requirement of high
signal-to-noise data and spatial resolution that have not been
obtained in the existing observations.

Recently Stawarz et al.(2008) presented a spectral emission model for
GPS sources (see also Ostorero et al. in this Proceedings). The main
contribution to the spectral energy distribution comes from the radio
lobes and hot spots of the $<1$~kpc size radio source embedded in the
radiation field of the host galaxy.  The X-ray emission is due to
Inverse Compton scattering off the relativistic electrons within the
radio source.  The UV-disk photons and IR-dust photons provide the
external radiation field to the hot spots, lobes and jet.  This field
dominates in smaller sources, while the synchroton photon field
intrinsic to the lobes dominates in larger sources, and GPS/CSS
sources exceeding 1~kpc in size will be dominated by the synchrotron
self-Compton emission process in the X-ray band. The predicted X-ray
luminosity depends on a few model parameters such as a jet power,
photon fields and the density of the ISM.  Note that this model
relates the X-ray emission to the radio source components while the
simulations described above consider the emission from the hot gas
heated by the radio source.

Both type of the X-ray emission, thermal and non-\break thermal, will
be present in GPS sources. Hot thermal gas will produce an X-ray
spectrum with emission lines, while the non-thermal continuum does not
have any spectral features. High resolution X-ray spectra are needed
in order to disentangle these two emission components.

\section{X-ray Observations of GPS/CSS Sources}

Before {\it Chandra} and XMM-{\it Newton} there was only one GPS
galaxy detected in X-rays by ASCA (O'Dea et al. 2000) and several GPS
quasars detected with HEAO-1 and ROSAT (see O'Dea 1998 for a
summary). During the last decade there have been several efforts to
increase the number of X-ray detections. Guainazzi et al. (2004, 2006)
and Vink et al. (2006) present X-ray samples of GPS galaxies,
Siemiginowska et al. (2003, 2005, 2008) discuss X-ray emission in
GPS/CSS quasars. Worrall et al.(2004) and O'Dea et al. (2006) study
details of the X-ray emission in a single GPS source.

{\it Chandra} and XMM-{\it Newton} observational capabilities are
complementary. {\it Chandra} exceptional point spread function (PSF)
allows for studies of X-ray morphology with $\sim1$~arcsec resolution
and the highest dynamic range available in the X-ray band today (Van
Speybroeck et al. 1997, Weisskopf et al. 2000). Because of its low
background it gives efficient detections of very faint sources and
relatively good quality spectra within 0.5-7~keV.  The PSF of the
XMM-{\it Newton} is too large for studies of the X-ray morphology of
GPS sources (Kirsch et al. 2004), but a very high effective area of
the telescope gives high quality spectra within 0.5-10~keV energy
range.

\begin{figure}
\includegraphics[width=40mm]{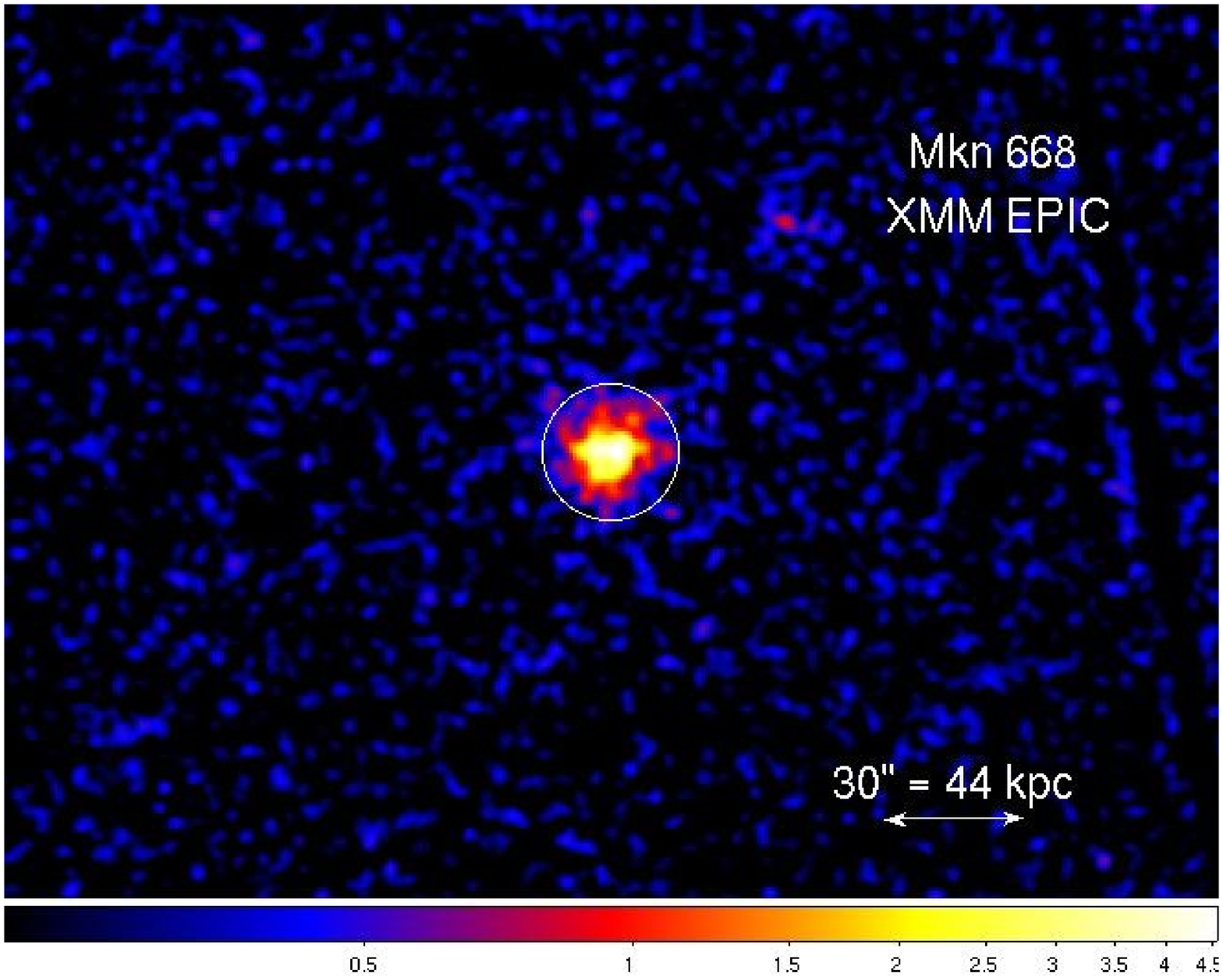}
\includegraphics[width=40mm]{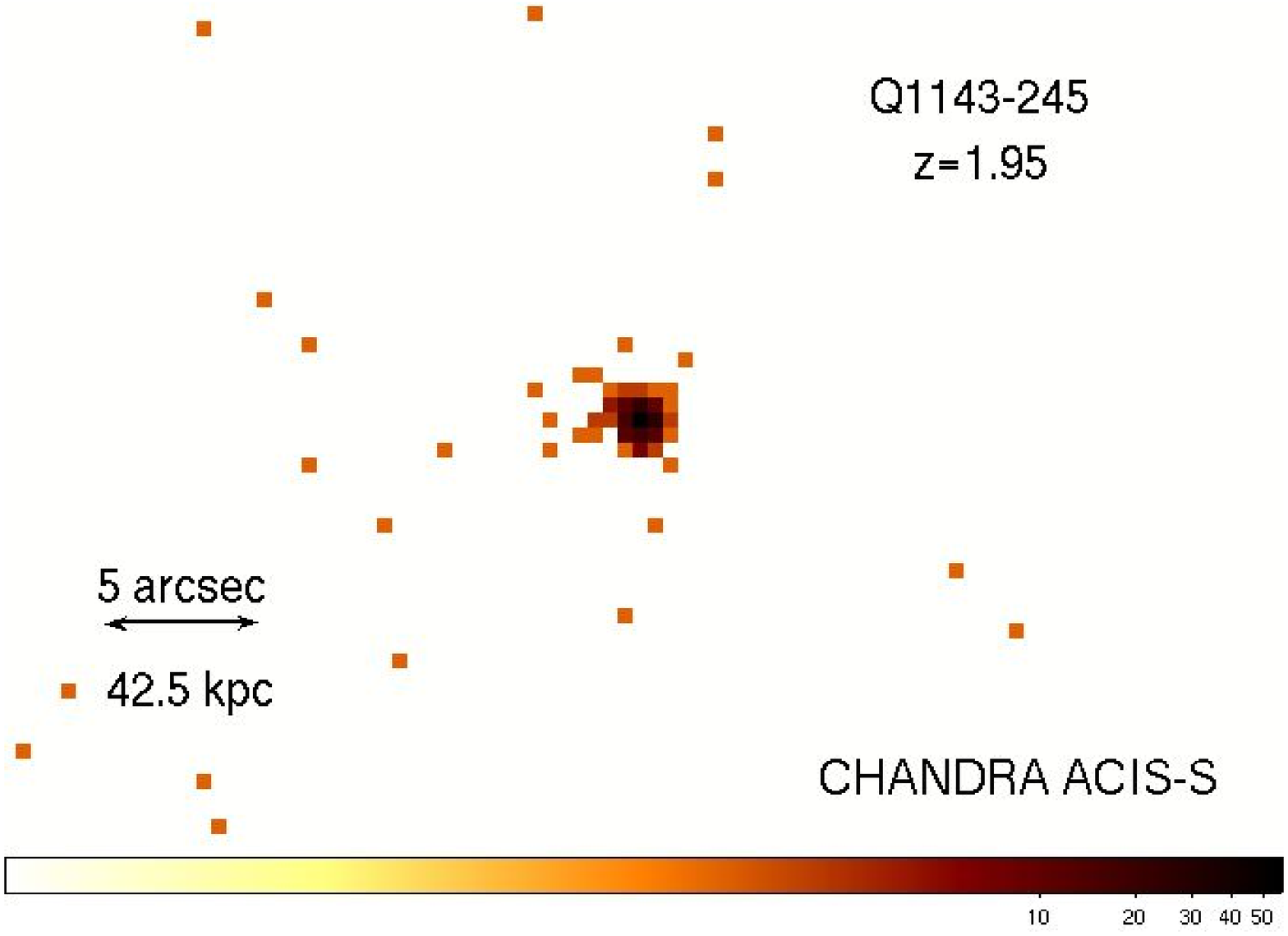}
\caption{{\bf Left:} XMM-{\it Newton} EPIC X-ray image of Mkn668 (z=0.076). 
A 30\arcsec\ radius circle is marked around the source. The source is
unresolved on this scale. A 30\arcsec\ scale bar equivalent to 44~kpc
is shown in the right corner. {\bf Right:} {\it Chandra} ACIS-S image
of Q1143-245 at z=1.95. 5\arcsec\ scale corresponding to 42.6~kpc at
the source redshift is marked in the left corner. Note that the radio
source size is much smaller than the resolution of these two
instruments.}
\label{fig1}
\end{figure}

\begin{figure}
\includegraphics[width=\columnwidth]{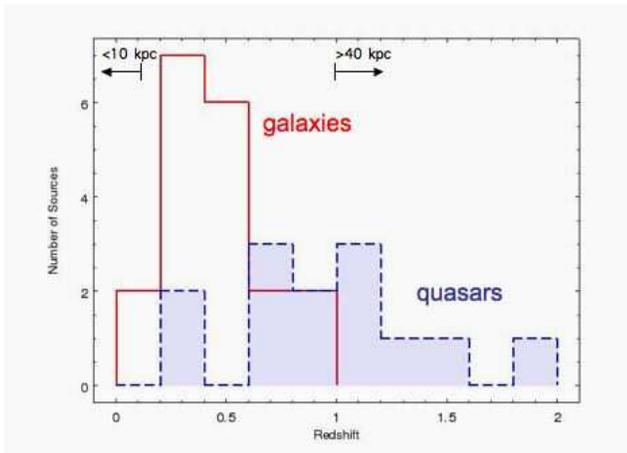}
\caption{Redshift distribution of GPS and CSS sources observed in X-rays. 
The histogram includes both {\it Chandra} and XMM-{\it Newton}
observations and consists of 19 galaxies and 13 quasars. The galaxies
are marked by a solid red line and quasars by a dashed blue line and a
shaded region. 5\arcsec scale corresponding to size $<10$~kpc and $>40$~kpc 
is marked by arrows. The data points are from Guainazzi et al. (2004, 2006),
Vink et al. (2006) and Siemiginowska et al. (2008).}
\label{fig2}
\end{figure}

The spatial resolution of modern X-ray telescopes is still much worse
than the resolution obtained by radio observations.  Figure~\ref{fig1}
shows the X-ray image of a nearby GPS source Mkn~668 at z=0.076 (1~arcsec =
1.46~kpc). A double radio structure ($< 10$~pc) is fully enclosed
within the X-ray point source observed with XMM-{\it Newton} EPIC
camera.  The highest redshfit GPS source in the X-ray sample is
plotted in Figure~\ref{fig2}. Q1143-245 (z=1.95) is an unresolved
X-ray point source in the {\it Chandra} ACIS-S observation and
1.5\arcsec corresponds to $\sim 12$~kpc at the quasar distance.

Table~\ref{tab1} lists the physical size in kpc for a source located at
different redshifts.  1~arcsec corresponds to a physical size of
179~parsec in Cen A, while it corresponds to 8~kpc for a source at
redshift 1.  The redshift distribution for a sample of sources
observed so far in X-rays is plotted in Figure~\ref{fig2}. The peak
distribution of the GPS galaxies is at $z\sim0.3-0.4$ and for a radio
source smaller than $< 10$~kpc cannot be resolved with the current
X-ray telescopes. For the GPS quasars the distribution spans evenly
the redshift between $0.4-2$. At redshift $z>1$ the 5~arcsec size of a
point source contains a source exceeding $>40$~kpc in size. GPS radio
structure is smaller than $<1$~kpc, CSS $<10$~kpc, so any associated
X-ray emission (as in Stawarz et al. 2008 model) would not be
resolved.


\begin{table}
\begin{center}
\caption{Physical Size vs. Redshfit}
\label{tab1}
\begin{tabular}{cccc}\hline
Redshift & 1\arcsec   & 5 \arcsec & Example\\ 
	 & kpc &  kpc & GPS Source\\
\hline
\\
0.00087 & 0.0179 & 0.0895   & Cen A \\
0.01 & 0.202 & 1 & \\
0.076 & 1.46 & 7.3 & Mkn~668 \\
0.1   & 1.8 & 9 & \\
0.668 & 7   & 35 & 0108+338 \\
1 & 8 & 40 7 \\
\hline
\end{tabular}
\end{center}
\end{table}

An unresolved X-ray point source will contain a source smaller than
$<3$\arcsec\, in {\it Chandra} and $<15$\arcsec\, in XMM{\it
-Newton} observations. There, in most GPS/CSS we can only study X-ray
spectral properties and may be able to disentangle different emission
components, evaluate absorption properties and intrinsic source
luminosity.  However, an X-ray emission on scales larger than
the PSF size can be studied with {\it Chandra} in nearby sources (see
Fig.~\ref{fig2}), e.g. a diffuse emission associated with cocoon,
Narrow Emission Line regions, super winds, X-ray cluster or relic
emission. In addition larger scale jets associated with the radio
source could be detected if their length exceeds $>3$\arcsec.

\subsection{Compact Source: X-ray Spectra}

GPS sources are faint in X-rays and have varying quality of the X-ray
spectrum, from a detection of a source with just a few counts in a
short $\sim5$~ksec {\it Chandra} exposure to a relatively good
spectra with $>1000$ counts in a longer 20-30~ksec XMM-{\it Newton}
observations.  Most spectra do not show any strong emission lines,
except for Mkn~668 (Guainazzi et al.2004), and can be characterized by
an absorbed power law model. Such a model provides an estimate of an
X-ray flux and probes the X-ray absorption properties.

Figure~3 shows a distribution of the intrinsic absorption column in
GPS galaxies observed so far in X-rays. There is a peak at
$\sim10^{22}$~atoms~cm$^{-2}$ in the distribution and most of the
galaxies are absorbed in X-rays. On the other hand the quasars show no
intrinsic absorption with detection limits below
$<10^{21}$~atoms~cm$^{-2}$ (Siemiginowska et al. 2008). The GPS
galaxies show an anticorrelation between the X-ray absorption column
and the radio source size (Guainazzi et al. 2006, Vink et al. 2006)
similar to the one noticed in the observations of HI 21~cm absorption
by Pihlstr{\"o}em, Conway \& Vermeulen (2003).  This may indicate an
evolutionary phase in radio source growth in which the initially
enshrouded nucleus is being ``clear out'' by an expanding radio
source.

\begin{figure}
\begin{center}
\includegraphics[width=70mm]{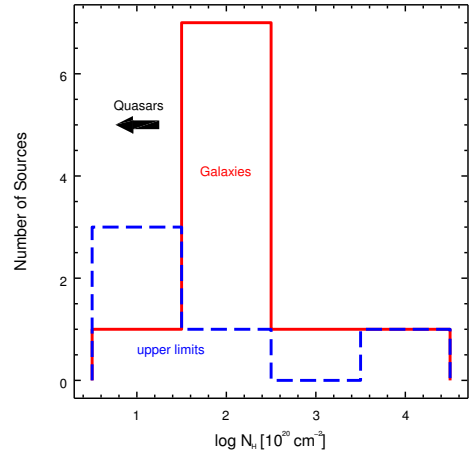}
\caption{Distribution of the intrinsic absorption column in GPS galaxies.
The histogram includes both {\it Chandra} and XMM-{\it Newton}
observations from Guainazzi et al. (2004, 2006), Vink et al. (2006),
Siemiginowska et al.2008, Tengstrand et al. (2008). Detections are
plotted with the solid red curve and upper limits with the dashed blue
curve.  The observed columns in GPS quasars reported by Siemiginowska
et al. (2008) are marked by a big arrow and all are below $N_H
<10^{21}$~atoms~cm$^{-2}$.}
\end{center}
\label{fig3}
\end{figure}

Given the X-ray spectra and the observed absorbing column one can
determine the source intrinsic X-ray luminosity. The new observations
indicate that the GPS galaxies are not X-ray quiet. Their X-ray
luminosity corrected for absorption exceeds $10^{42}$~ergs~s$^{-1}$
and matches the X-ray luminosities of GPS quasars and FRII type radio
galaxies (Guainazzi et al.2006, Tengstrand et al. 2008).

\begin{figure}
\includegraphics[width=50mm, angle=90]{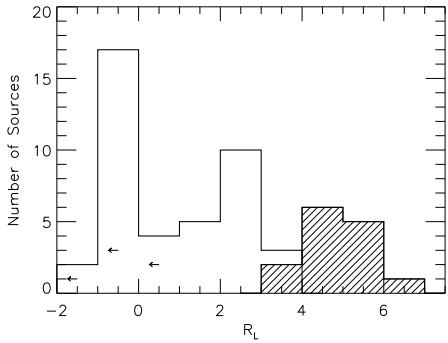}
\caption{Radio loudness of the GPS quasars from Siemiginowska et al.(2008) 
compared to  the other radio-loudquasars from Elvis et al.(1994).
GPS/CSS quasars are marked by shaded histogram and they are more radio
loud.}
\label{fig4}
\end{figure}

\begin{figure}
\includegraphics[width=50mm, angle=90]{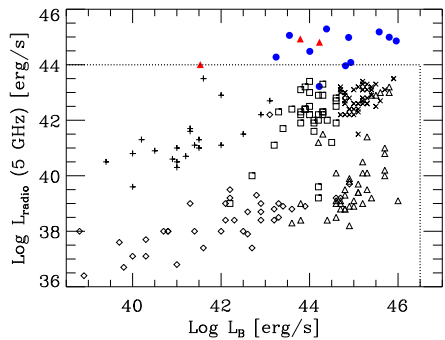}
\caption{Radio loudness in comparison to the B-band luminosity for a different 
classes of sources in Sikora et al. (2007). The location of the GPS quasars
(circle) and galaxies (triangle) is marked on the original diagram of
Sikora et al. Most GPS sources are more radio loud, optically brighter
and located outside a range (marked by a dashed line) observed in the
other sources. Note B-luminosity for GPS galaxies is typically
undetermined, because of the intrinsic absorption of the nucleus. }
\label{fig5} 
\end{figure}

With the new observations we can also study the spectral energy
distribution of the GPS sources from radio to X-rays. Such analysis
indicate that the GPS/CSS quasars are more radio loud than the other
radio loud quasars (see Fig.~\ref{fig4}), where the radio loudness is
defined as a ratio between the radio and optical (B band)
luminosity. A comparison with a larger sample of variety of radio loud
objects in Sikora, Stawarz \& Lasota (2007) shows that the GPS quasars
and galaxies are most radio luminous for a given optical luminosity
(see Fig.~\ref{fig5}). This may suggest a higher radiative efficiency
in a younger (smaller) GPS source as postulated by the theory (see
Begelman, 1997).

We can summarize the main observational results based on the X-ray spectra: 

\begin{itemize}

\item[$\bullet$] Absorption: 
(1) high ($>10^{22}$ cm$^{-2}$) 
column densities are common in galaxies; 
(2) there is a
tentative dependence of the column density on the distance between the
hot spots/lobes X-ray; 
(3) X-ray column density is always larger than the HI
column observed in radio.

\item[$\bullet$]  Intrinsic X-ray Luminosity:
(1) GPS galaxies are not X-ray quiet; 
(2) GPS/CSS quasars show an indication that they are X-ray weak.

\item[$\bullet$] X-ray loudness: 
(1) X-ray to radio luminosity ratios are similar to the ratios
observed in FRI galaxies; \break (2) GPS/CSS quasars are the most
radio loud objects.

\end{itemize}

\subsection{X-ray Morphology (on kpc scales)}

GPS radio structure is much smaller than the spatial resolution
capabilities of the current X-ray instruments, so the GPS source
remains unresolved (even in the lowest redshift sources) in all X-ray
observations to date. X-ray morphology on scales exceeding the GPS
source size can be studied with {\it Chandra}.  Several types of X-ray
morphology have been detected in the GPS and CSS sources:

\begin{itemize}

\item[$\bullet$] X-ray jets were detected on large scales up to 
hundreds kpc away from the GPS source: PKS1127-145 (Siemiginowska et
al. 2002, 2007), B2~0738+393 (Fig.6, \break Siemiginowska et al. 2003).

\item[$\bullet$] an irregularly shaped diffuse X-ray emission:\break
PKS~B1345+125 (Siemiginowska et al. 2008,
Guainazzi et al. in preparation)

\item[$\bullet$] a secondary X-ray emission associated with a radio \break 
source located 20\arcsec to the west of a double nucleus GPS galaxy 
PKS 0941-080  (Siemiginowska et al. 2008)

\item[$\bullet$] an X-ray emission associated with a cluster of galaxies 
at the GPS source redshift: CSS quasar 3C186 is in the center of an
X-ray cluster (Siemiginowska et al. 2005)

\end{itemize}

\begin{figure}
\begin{center}
\includegraphics[width=50mm]{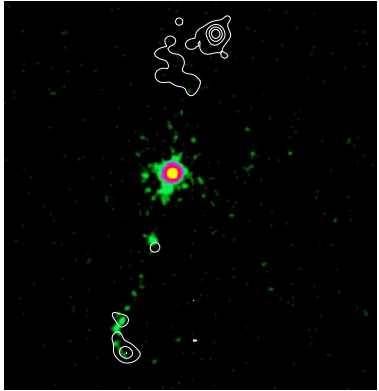}
\caption{A smoothed {\it Chandra} ACIS-S (0.3-7~keV) image of B2~0738+393 
overlayed with the radio contours (VLA-1.4~GHz).  The X-ray jet is
narrow, curves and follows the radio structure to the south. The jet
ends with a hot spot at the southernmost part of the radio lobe. A
knot at $\sim 13$ arcsec away from the core and its emission is
consistent with the X-rays being created by the inverse Compton
scattering of the cosmic microwave background (CMB) photons and
requires jet bulk Lorentz factors of a few ($\Gamma_{bulk} \sim 5-7$)
(see Siemiginowska et al. (2003) for details).}
\end{center}
\label{0738}
\end{figure}

A large scale X-ray emission associated with the GPS source may also
indicate intermittency of a central AGN. Baum et al. (1990) postulated
that an extended radio emission detected $\sim 20\arcsec$ away from
the GPS galaxy 0108+388 is a relic of an earlier active phase of the
source. Reynolds \& Begelman (1997) argued that intermittent activity
can explain the observed statistics of GPS sources.  Intermittency or
``modulated jet activity'' has been considered as a mechanism shaping
the morphology of large scale jets (for details see Stawarz et
al. 2004 and examples: 3C273 by Stawarz 2004, and Jester et al. 2006;
PKS1127-145 Siemiginowska et al. 2007). The newborn portion of the jet
resembles the GPS source, but at later time it will be observed as a
knot in the extended jet.

A large scale emission allows also for studies of interactions between
the radio source and the environment.  A detection of an X-ray cluster
in the {\it Chandra} observation of 3C186 (Fig.7) provided a way to
study the surroundings of the radio source. The measured temperature
and density of the X-ray emitting hot cluster gas allow for an
estimate of the gas pressure. This in turn could be compared to the
pressure measured in the radio lobes. In 3C186 the pressure in the
radio lobes exceeded the thermal gas pressure by 2 orders of magnitude
(Siemiginowska et al.2005).  This observation therefore allowed to
conclude that the jet is not frustrated and the radio source expands
with no disruptions.

\begin{figure}
\begin{center}
\includegraphics[width=60mm]{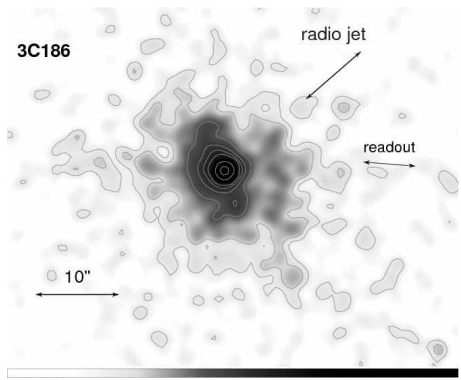}
\caption{Smoothed {\it Chandra} ACIS-S (0.3-7~keV) image of 3C~186
The diffuse cluster emission extends to $\sim$120~kpc from the central
quasar. The 10 arcsec (82~kpc) size is marked.  The direction of an
unresolved 2~arcsec radio jet is also marked.}
\end{center}
\label{3c186}
\end{figure}

Fig.8 shows smoothed {\it Chandra} ACIS-S images of \break
PKS~B1345+125 in the soft (0.5-2~keV) and hard (2-10~keV) bands. The
diffuse X-ray emission seen in the soft band may originate as a
thermal emission associated with the galaxy halo. Its size agrees with
the size of the extended optical line region studied kinematically in
the optical and described by Holt, Tadhunter \& Morganti (2003). The
X-ray emission is also stretched towards the South-West similarly to
the optical emission and is aligned also agrees with the VLBI jet axis
(Stanghellini et al. 2001) suggesting that it may be related to the
expanding GPS source. More detailed discussion is presented in
Guainazzi et al. (2008 in preparation).

Interestingly there is so far no detection of a hot (kT$\sim$~ a few
keV) cocoon that is predicted by theoretical models of the expanding
radio source.  Is this because we did not observe the right targets or
we are limited by the spatial resolution of the telescopes.  O'Dea et
al. (2006) reported a thermal component in an XMM-Newton spectrum of
3C303.1 (z=0.267). They suggest that this thermal emission could be
due to plasma heated by the shock of the expanding radio source. The
source is unresolved in this X-ray observation and only the spectrum
could be used to disentangle the emission components. Future {\it
Chandra} observation may allow for more detailed investigations of the
X-ray properties on a few arcsec scales.

\begin{figure}
\begin{center}
\includegraphics[width=70mm]{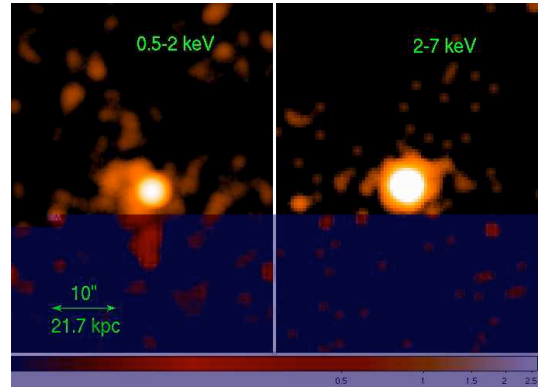}
\caption{Smoothed {\it Chandra} ACIS-S images of PKS~1345+125. 
{\bf Left:} The soft (0.5-2~keV) band showing the extended emission surrounding the core. Them 10\arcsec scale is marked at the left corner and it is the same in both images.
{\bf Right:} The hard(2-7~keV) band image. It is dominated by the
unresolved core.}
\end{center}
\label{pks1345}
\end{figure}

\section{Conclusions and Future Perspectives}

A number of pointed X-ray observations of GPS and CSS sources has
increased significantly over the last decade. \break However, a
complete sample is still not available and there is therefore no
statistically solid and systematic study of the X-ray properties of
such a sample.  The current observed trends need to be confirmed in
the future.  We are continuing our efforts to obtain X-ray
observations of all the GPS sources in Stanghellini et al. (1998) and
making a slow \break progress.

There are many remaining questions as we do not know answers to the
observed correlations, the processes dominating X-ray emission, radio
source environment. The basic questions about the nature of the
activity is still not answered: Is the radio source intermittent? Is
the source short lived? Does the GPS radio structure fade away? Many
questions that have been asked at this workshop were related to the
source nature and evolution. They have to be investigated using
multi-wavelength data. Samples that cover ra-
dio-optical-UV-X-rays-$\gamma$-rays are needed to test models and
understand source nature and connection to a population of large scale
radio sources.

There are plans for future X-rays and $\gamma$-rays missions that will
become active and provide new information about the high energy
processes in the GPS sources. FERMI \break (GLAST) has been
successfully launched this year. This is the first $\gamma$-ray
instrument capable of detecting GPS sources if they emit
$\gamma$-rays. FERMI will collect the data at $\gamma$-ray energies
$>10^{22}$~Hz) that may provide tests and possible constraints on the
emission processes from the GPS radio lobes predicted by Stawarz et al
(2008).

New high energy X-ray missions with a large collective area such as
NuStar (Harrison et al. 2005, Koglin et al. 2005) and EXIST (Grindlay
2007) are being built and will become available in the near
future. They will be capable of detecting X-ray emission from GPS/CSS
sources and will provide low resolution X-ray spectra at energies
$>$10~keV for the first time.  The high resolution X-ray spectra will
become available in the next decade when the X-ray International
Observatory (IXO) is placed on orbit. This new mission will also have
good imaging capabilities. However, the highest resolution X-ray
spectra will allow for detailed studies of the absorption and emission
line structure, so will give information about properties of the gas
intrinsic to the source and also of a thermal emission components. For
example we will be able to distinguish between the shock heated plasma
emission and the synchrotron non-thermal emission in 3C303.1 presented
by O'Dea (2008) at this workshop.

\medskip

\acknowledgements

SOC and LOC members are acknowledged for organization of a
great meeting.   
This research is funded in part by NASA contract NAS8-39073 and grant
NNX07AQ55G. Partial support for this work was provided by the National
Aeronautics and Space Administration through Chandra Award Number
GO5-6113X, GO7-8103X-R issued by the Chandra X-Ray Observatory Center,
which is operated by the Smithsonian Astrophysical Observatory for and
on behalf of NASA under contract NAS8-39073.

\end{document}